\documentclass[prb, reprint, 9pt, superscriptaddress,
notitlepage, nofootinbib, longbibliography,
floatfix]{revtex4-2}
\usepackage{natbib}
\usepackage{graphicx}
\usepackage{epsfig}
\usepackage{amsfonts} 
\usepackage{amsmath} 
\usepackage{amssymb}
\usepackage{diagbox}

\usepackage{bm}
\usepackage{braket}
\usepackage{color}
\usepackage{physics} 
\usepackage{bbm}
\usepackage{hyperref}
\usepackage{subcaption}

\renewcommand{\raggedright}{\leftskip=0pt \rightskip=0pt plus 0cm}
\newcommand{\be}{\begin{equation}}
\newcommand{\ee}{\end{equation}}
\newcommand{\ba}{\begin{eqnarray}}
\newcommand{\ea}{\end{eqnarray}}

\newcommand{\mA}{\mathcal{A}}
\newcommand{\mB}{\mathcal{B}}

\definecolor{LinkColor}{rgb}{0.256,0.439,0.588}
\hypersetup{
	colorlinks=true,
	citecolor=LinkColor,
	linkcolor=LinkColor,
	urlcolor=LinkColor
}

\usepackage{xcolor}

\definecolor{gr}{rgb}{0,0,0}

\begin{document}

\title{Entanglement cost in topological stabilizer models at finite temperature}
\author{Tsung-Cheng Lu}
\affiliation{Perimeter Institute for Theoretical Physics, Waterloo, Ontario N2L 2Y5, Canada}
\author{En-Jui Kuo}
\affiliation{Department of Physics, University of Maryland, College Park, MD 20742, USA}
\affiliation{Joint Quantum Institute, NIST/University of Maryland, College Park, MD 20742, USA}
\author{Hung-Hwa Lin}
\affiliation{Department of Physics, University of California at San Diego, La Jolla, California 92093, USA}

\begin{abstract}
The notion of entanglement has been useful for characterizing universal properties of quantum phases of matter. From the perspective of quantum information theory, it is tempting to ask whether their entanglement structures possess any operational meanings, e.g., quantifying the cost of preparing an entangled system via free operations such as the local operations and classical communication (LOCC). While the answer is affirmative for pure states in that entanglement entropy coincides with entanglement cost, the case for mixed states is less understood. To this end, we study the entanglement cost required to prepare the thermal Gibbs states of certain many-body systems under positive-partial-transpose (PPT) preserving operations, a set of free operations that include LOCC. Specifically, we show that for the Gibbs states of $d$-dimensional toric code models for $d = 2, 3, 4$,  the PPT entanglement cost exactly equals entanglement negativity, a measure of mixed-state entanglement that has been known to diagnose topological order at finite temperature. 

\end{abstract}

\maketitle



\section{Introduction}

The concept of entanglement has been powerful for revealing universal properties of quantum phases of matter. A well-known example is the topologically ordered states of matter, which possess long-range entanglement that is irrespective of any microscopic details of the systems\cite{Kitaev_topo_entanglement_2006,levin_wen_entanglement_2006}. On the other hand, from the perspective of quantum information theory, one often regards entanglement as a resource for tasks that are impossible/difficult to achieve by means of classical resources\cite{entanglement_2009_Horodecki}, and therefore, it is crucial to characterize entanglement in an operationally meaningful manner. This motivates us to ask: whether the entanglement structure of many-body systems has any operational meanings for protocols relevant to quantum information.

There are two primary notions for characterizing the operational meaning of  entanglement: distillable entanglement and entanglement cost\cite{cost_distillable_pure_1996,distill_entanglement_1996,Wootters_mixed_state_1996}. The former quantifies the largest rate at which maximally entangled states can be distilled from a given quantum state using LOCC, i.e. local operations and classical communication. The latter measures the smallest rate at which maximally entangled states are required to prepare a target state using LOCC. While for pure state, the most common entanglement measure, namely entanglement entropy, already measures both the distillable entanglement and entanglement  cost\cite{cost_distillable_pure_1996}, it is generally difficult to characterize either of these two quantities in mixed states. For instance, (asymptotic) LOCC entanglement cost has been proven to be the regularized entanglement of formation\cite{2001_cost_terhal}, but such a quantity cannot be efficiently computed in many-body systems. On the other hand, the only known mixed-state entanglement measure that is simple to compute is entanglement negativity (also dubbed logarithmic negativity)\cite{peres_1996,Horodecki_1996,Eisert_negativity_1999,werner_vidal_2002}, but in general it has no operational meaning, albeit being an upper bound on distillable entanglement and teleportation capacity\cite{werner_vidal_2002}. As such, it has been a long-standing question for finding a computable mixed-state entanglement measure endowed with an operational meaning.

Progress has been made by enlarging the set of free operations from LOCC to positive-partial-transpose (PPT) preserving operations, whose mathematical structure is much easier to characterize compared to LOCC\cite{chitambar2014everything}. A defining feature of these PPT-preserving operations $\Lambda$ is that they transform a state $\hat{\rho}$ whose partial transpose is positive to another state whose partial transpose remains positive\cite{rains1999bound,rains1999rigorous}, namely, $\hat{\rho}^{\Gamma} \geq 0 \Rightarrow  \left[\Lambda \left( \hat{\rho} \right) \right]^{\Gamma} \geq 0$. Here $\hat{\rho}^\Gamma$ denotes the partial transpose of $\hat{\rho}$, i.e. $\hat{\rho}^{\Gamma} = \sum_{a,b,a',b'}  \rho_{ a,b;a',b'   } \ket{ a',b } \bra{ a,b'}  $ given a matrix representation of $\hat{\rho}$ acting on a bipartite Hilbert space $\mathcal{ H}_{\mathcal{A}} \otimes \mathcal{ H}_{\mathcal{B}} $:  $\hat{\rho}=\sum_{a,b,a',b'}  \rho_{ a,b;a',b'   } \ket{ a,b } \bra{ a',b'}  $. PPT-preserving operations are more powerful than LOCC as they allow to create bound entangled states whose entanglement cannot be distilled under LOCC\cite{bound_entanglement_horodecki_1998}. By considering this enlarged set of free operations, Audenaert, Plenio, and Eisert studied the entanglement cost under PPT-preserving operations\cite{audenaert2002entanglement}, and provided a lower bound and an upper bound for PPT entanglement cost. In particular, their bounds imply that the PPT entanglement cost exactly equals entanglement negativity for any states $\hat{\rho}$ satisfying $\abs{\hat{\rho}^{\Gamma}}^{\Gamma} \geq 0$, where the absolute value sign acts as $\abs{\hat{\rho}^{\Gamma}   }=\sqrt{ \hat{\rho}^{\Gamma}  (\hat{\rho}^{\Gamma} )^{\dagger} }$. While there exists no systematic characterization for states that fulfill such a condition, it holds for  pure states, two-qubits states, bosonic Gaussian states, and Werner states \cite{audenaert2002asymptotic,audenaert2002entanglement,Ishizaka_binegativity_2004}, and therefore entanglement negativity is an operationally meaningful quantity for these classes of states.

%
%

Motivated by the result in Ref.\cite{audenaert2002entanglement}, we will study the PPT entanglement cost for a class of states relevant to topologically ordered phases of matter\cite{wen1989vacuum,wen1990ground,wen1990topological}, namely, the toric code model in $d$ space dimensions\cite{kitave_2d_2003,dennis2002}. Toric code exhibits a topological order at zero temperature that is robust under any weak local perturbations. As such, it allows for a robust encoding of qubits in its ground subspace. Moreover, the topological order of toric code in 4 space dimensions can survive thermal fluctuations, and therefore provides a genuine stable, self-correcting quantum memory and robust topological order below a certain non-zero critical temperature\cite{dennis2002,hastings2011,yoshida2011}. As discussed in Ref.\cite{lu2020detecting,lu_vijay_spt}, the non-trivial topological order at finite temperature can be diagnosed by a universal, long-range component of entanglement negativity. This motivates us to explore the operational meaning of entanglement negativity in this class of models.

By focusing on the toric code models described by a Gibbs state $\hat{\rho}\sim e^{ -\beta \hat{H}}$ in various space dimensions, we show that their Gibbs states satisfy $\abs{\hat{\rho}^{\Gamma}}^{\Gamma} \geq 0$ for any temperature, therefore indicating that entanglement negativity exactly equals PPT entanglement cost. Our work therefore provides a notable class of examples for non-trivial many-body quantum states whose entanglement cost is tractable. In particular, we develop a formalism for computing the spectrum of $\abs{\hat{\rho}^{\Gamma}}^{\Gamma}$, which we name ``binegativity spectrum''\footnote{The term ``binegativity spectrum'' is motivated by Ref.\cite{audenaert2002entanglement}, where the matrix $\abs{\hat{\rho}^{\Gamma}}^{\Gamma}$ is called binegativity.}, for any Gibbs states of stabilizer Hamiltonians.

\section{PPT entanglement cost and binegativity spectrum}

A quantum channel $\Lambda_{A_0B_0\to AB}$ is a completely positive trace-preserving map from the input systems $A_0$, $B_0$ to output systems $A$, $B$, where one party Alice possesses $A_0$ and $A$ and another party Bob possesses $B_0$ and $B$. $\Lambda_{A_0B_0\to AB}$ is positive-partial-transpose (PPT) preserving if and only if $T_B\circ \Lambda_{A_0B_0\to AB} \circ T_{B_0}$ is completely positive\cite{rains1999bound,rains1999rigorous}, where $T_{B_0}$ and $T_B$ is the partial transposition acting on the subsystem $B_0$ of the input state and the  subsystem $B$ of the output state.  This implies  if the partial transpose of an input state is positive, then the partial transpose of the output state generated by the map $\Lambda$ will be positive as well. 

Based on the notion of the PPT-preserving channel, one can  define the corresponding PPT entanglement cost. First, we consider a maximally entangled pure state $\Phi_{A_0B_0}^{d}= \ket{\Phi}\bra{\Phi}$, where $\ket{\Phi}  =\frac{1}{\sqrt{d}}  \sum_{i=1}^d \ket{i}_{A_0}\ket{i}_{B_0}$ with $\{  \ket{ i}_{A_0 } \}$ and $\{   \ket{ i }_{  B_0}    \}$ being the orthogonal bases in $A_0$ and $B_0$.  Treating such a state as a resource, one asks how much entanglement is required to generate a target state $\hat{\rho}_{AB}$ acting on the bipartite Hilbert space $\mathcal{H}= \mathcal{H}_A \otimes \mathcal{H}_B$ using PPT-preserving channels. This question motivates to define the one-shot PPT exact entanglement cost  $E_{\text{PPT}}^{\left(1\right)}\left(\hat{\rho}_{AB}\right)$ as the logarithm of the minimum Schmidt rank of a maximally entangled state minimized over the  PPT-preserving channels used to prepare $\hat{\rho}_{AB}$: 
\begin{equation}
	\begin{split}
&	E_{\text{PPT}}^{\left(1\right)}\left(\hat{\rho}_{AB}\right)\equiv\\
 & \inf_{d\in\mathbb{N},\Lambda\in\text{PPT}}\left\{ \log_2 d:\hat{\rho}_{AB}=\Lambda_{A_0B_0\to AB}\left(\Phi_{A_0 B_0}^{d}\right)\right\} \label{eq:E_ppt_oneshot}, 
	\end{split}
\end{equation}
It follows that one can define the (asymtopic) PPT entanglement cost $E_{\text{PPT}}\left(\hat{\rho}_{AB}\right)$ as the average cost of preparing infinite copies of $\hat{\rho}_{AB}$\cite{audenaert2002entanglement}
\begin{align}
	E_{\text{PPT}}\left(\hat{\rho}_{AB}\right)\equiv \lim_{n\to\infty}\inf\frac{1}{n}E_{\text{PPT}}^{\left(1\right)}\left(\hat{\rho}_{AB}^{\otimes n}\right)\,.\label{eq:E_ppt_kappa}
\end{align}

As shown in Ref.\cite{audenaert2002entanglement}, PPT entanglement cost satisfies the following inequality
\begin{align}\label{eq:kappa_bound}
E_{N}\left(\hat{\rho}\right)\leq & E_{\textrm{PPT}}     \left(\hat{\rho}\right)\leq\log Z\left(\hat{\rho}\right)\,,
\end{align}
where $E_N(\hat{\rho} ) $ is entanglement negativity defined as $E_N(\hat{\rho})  = \log \left( \tr \abs{\hat{\rho}^{\Gamma}} \right)$\cite{peres_1996,Horodecki_1996,Eisert_negativity_1999,werner_vidal_2002}, and 

\begin{align}
	Z\left(\hat{\rho}\right)= & \text{Tr}\left|\hat{\rho}^{\Gamma}\right|+\dim\left(\hat{\rho}\right)\max\left(0,-\lambda_{\text{min}}\left(\left|\hat{\rho}^{\Gamma}\right|^{\Gamma}\right)\right)\,.
\end{align}
with $\lambda_{\text{min}}\left(\left|\hat{\rho}^{\Gamma}\right|^{\Gamma}\right) $ being the minimal eigenvalue in the binegativity spectrum, i.e. the eigenspectrum of $\abs{\hat{\rho}^{\Gamma}}^{\Gamma}$. Crucially, if the binegativity spectrum is non-negative, the lower and upper bound in Eq.\ref{eq:kappa_bound} coincide,  indicating the equivalence between PPT entanglement cost and  entanglement negativity: 

\begin{equation}
E_{\textrm{PPT}}  (\hat{\rho}) = E_N(\hat{\rho}).  
\end{equation}

\section{Binegativity spectrum in toric code models at finite temperature}
As discussed above, when a state has non-negative binegativity spectrum, i.e. $\abs{\rho^{\Gamma}}^\Gamma \geq 0$ , PPT entanglement cost is simply given by entanglement negativity. Below we will present a general formalism for computing the binegativity spectrum for Gibbs states of stabilizer models, and utilizing it to show that the Gibbs state of $d$-dimensional toric code model for $d=2,3,4$ satisfies $\abs{\rho^{\Gamma}}^\Gamma \geq 0$. Note that in the discussion, operators will be denoted with a hat to distinguish them from classical numbers. 

\subsection{General formalism for stabilizer models}
Given a set of stabilizers $\{ \hat{\theta}_i \}$\cite{gottesman1997stabilizer}, where each stabilizer is a tensor product of Pauli operators acting on qubits, one can define the stabilizer Hamiltonian $\hat{H}  =  -  J\sum_i  \hat{\theta}_i$, and the corresponding thermal Gibbs state at the inverse temperature $\beta$ is $\hat{\rho}\sim  e^{ -\beta \hat{H} } =  e^{  \beta J\sum_i \hat{\theta}_i }  =  \prod_i e^{\beta J \hat{\theta }_i}$, where the last equality follows from the fact that all stabilizers commute with each other. Using the expansion $e^{ \beta J  \hat{\theta}_i  }  = \cosh(\beta J) \sum_{s_i=0,1} (t\hat{\theta}_i)^{s_i}  $ with $t =  \tanh( \beta J)$, the Gibbs state can be written as $\hat{\rho}\sim \sum_{\{s_i\}}  \prod_i \left( t \hat{\theta}_i\right)^{s_i} $. Assuming all stabilizers do not involve  Pauli-Ys (the cases with Pauli-Ys involved can be easily generalized, see Appendix.\ref{appendix:spectrum}), taking a partial transpose on the subregion $A$ introduces a sign $\psi(   \{s_i\})  \in \{  1, -1\}$: 
\begin{equation}
\hat{\rho}^{\Gamma} \sim  \sum_{\{s_i\}} \psi( \{s_i\}  )  \prod_i \left( t \hat{\theta}_i\right)^{s_i}  . 
\end{equation}

Here the sign $\psi(\{s_i \})$ is determined by the anticommutation relation among $\hat{\theta}_i$ in $\prod_i \hat{\theta}_i^{s_i}$ when restricted on the subregion $\mA$. To compute it, we introduce $\hat{\theta}_i|_{\mA}  $ as the operator in $\hat{\theta}_i$ supported on $\mA$, and introduce the matrix $C$ that encodes the commutation relation among these restricted stabilizers: $C_{ij}=0, 1$ for $[ \hat{\theta}_i|_{\mA}, \hat{\theta}_j|_{\mA} ] =0  $ and  $\{ \hat{\theta}_i|_{\mA}, \hat{\theta}_j|_{\mA} \} =1  $ respectively. One finds that the sign $\psi(\{s_i\}  ) $ is $1, -1$ when there is an even/odd number of pairs of restricted stabilizers $\hat{\theta}_i|_{\mA}$ that anticommute with each other, namely,  $\psi(\{s_i\}  ) = (-1)^{  \sum_{  i<j } s_iC_{ij}s_j    }$. Since $\hat{\theta}_j$ commute with each other, the negativity spectrum $\rho^{\Gamma}$, i.e. the eigenspectrum of $\hat{ \rho}^{\Gamma} $, can be obtained by replacing $\hat{\theta}_i$ with $\theta_i \in  \{    1,-1\}$:  

\begin{equation}
	\rho^{\Gamma}(T, \{\theta_i\}) \sim  \sum_{ \{s_i\}} \psi( \{s_i\}  )  \prod_i \left( t \theta_i\right)^{s_i}  . 
\end{equation}
The eigenspectrum allows us to express $\hat{\rho}^{\Gamma} $ in terms of the stabilizers $\{ \hat{\theta}_i  \}  $: 
\begin{equation}
	\hat{\rho}^{\Gamma }  = \sum_{  \{\theta_i\}  }  \rho^{\Gamma    }(T,  \{\theta_i\}  ) \prod_{i} \frac{1+  \theta_i \hat{\theta}_i   }{2},
\end{equation}
in which one can take the absolute value: $	\abs{\hat{\rho}^{\Gamma }  }= \sum_{   \{\theta_i\}    }  \abs{ \rho^{\Gamma    }( \{\theta_i\} )} \prod_{i} \frac{1+  \theta_i \hat{\theta}_i   }{2}$. To derive the binegativity spectrum, the eigenspectrum of $\abs{ \hat{\rho}^{\Gamma}   }^{\Gamma}$, we can expand the projector followed by taking a partial transpose and find $\abs{		\hat{\rho}^{\Gamma } }^{\Gamma} = \sum_{\{\theta_i\}}  \abs{\rho^{\Gamma    }(T, \{\theta_i\}   ) }   \left[ \sum_{\{\tau_i\}} \psi( \{ \tau_i \})    \prod_i  (\theta_i \hat{\theta}_i )^{\tau_i}      \right] $. Again, since stabilizers commute, we can replace $\hat{\theta}_{i}$ by $g_i\in \{1, -1\}$ to derive the binegativity spectrum: 

\begin{equation}\label{main:binegativity}
\abs{\rho^{\Gamma}   }^{\Gamma} ( T,\{  g_i \} )= 	\sum_{\{\theta_i \}  }  \abs{\rho^{\Gamma    }( T, \{  \theta_i \} )  } \rho^{\Gamma    }( T=0,  \{ \theta_i g_i \} ).
\end{equation}
Alternatively, by a change of variables, the above equation can be written as 
\begin{equation}\label{main:binegativity_v2}
	\abs{\rho^{\Gamma}   }^{\Gamma} (T,\{g_i\})= 	\sum_{\{\theta_i \}  }  \abs{\rho^{\Gamma    }( T, \{  \theta_ig_i \} )  } \rho^{\Gamma    }( T=0,  \{ \theta_i \} ).
\end{equation}
Therefore, the negativity spectrum completely determines the binegativity spectrum.

The result above can be further simplified  by decomposing a Gibbs state into a bulk part and a boundary part, namely,  $\hat{\rho}\sim \hat{\rho}_{  \textrm{bulk}  } ~  \hat{\rho}_{\partial}  $, where $\hat{\rho}_{ \textrm{bulk} }\sim e^{  \beta   J  \sum_{i  \in  \textrm{bulk} } \hat{\theta}_i    }$ contains the stabilizers only acting on region $\mA $ or $\mB$,  and $\hat{\rho}_{\partial }\sim e^{  \beta   J  \sum_{i  \in \partial } \hat{\theta}_i    }$ contains the stabilizers acting on both $\mA$ and $\mB$ (i.e. the bipartition boundary between $\mA$ and $\mB$). Since only those boundary stabilizers can anticommute when restricted in a subregion, the partial transpose only acts non-trivially on $\hat{\rho}_{\partial   }$, and the negativity spectrum can be factorized as $\rho^{\Gamma}  \sim \rho_{ \textrm{bulk}  } ~\rho^{\Gamma}_{\partial}  $. Plugging this decomposition into Eq.\ref{main:binegativity_v2} gives

\begin{equation}
\begin{split}
\abs{\rho^{\Gamma}   }^{\Gamma} (T,\{ g_i\}) = &\sum^{\textrm{bulk}}_{ \{   \theta_{i}  \} }  \sum^{\partial   }_{\{ \theta_i  \}  }  \abs{ \rho_{\textrm{bulk} }( T,  \{ \theta_ig_i   \}  )  \rho^{\Gamma}_{\partial} (T, \{\theta_i g_i \}  )     }  \\
&  \rho_{\textrm{bulk} }( T=0,  \{ \theta_i \}  )  \rho^{\Gamma}_{\partial} (T=0, \{\theta_i  \} ),   
\end{split}
\end{equation}
where $\sum^{\textrm{bulk}}_{ \{   \theta_{i}  \} } $ and $ \sum^{\partial   }_{\{ \theta_i  \}  } $ denote the summation over the stabilizers in the bulk and the stabilizers on the boundary. Using the fact that $ \rho_{\textrm{bulk} }( T=0,  \{ \theta_i   \}  ) $ projects to $\theta_i =1$ for all the bulk stabilizers and $ \rho_{\textrm{bulk} }( T,  \{ \theta_i  g_i \}  ) $ must be non-negative, the binegativity spectrum can be simplified as $
	\abs{\rho^{\Gamma}   }^{\Gamma} (T, \{g_i\}) $
	\begin{equation}
		\sim \rho_{ \textrm{bulk} }(T, \{g_i\}  ) 	\sum_{\{\theta_i \}  }^{\partial}  \abs{\rho_{\partial}^{\Gamma    }( T, \{  \theta_ig_i \} )  } \rho_{\partial}^{\Gamma    }( T=0,  \{ \theta_i \} ).
\end{equation}
Consequently, the sign of the binegativity spectrum of a thermal Gibbs state of stabilizer models is fully  determined by the boundary part of its negativity spectrum:
\begin{equation}\label{main:boundary}
	\abs{\rho^{\Gamma}}^{\Gamma}(T,\{g_i \})  \sim 	\sum_{\{\theta_i \}  }^{\partial}  \abs{\rho_{\partial}^{\Gamma    }( T, \{  \theta_ig_i \} )  } \rho_{\partial}^{\Gamma    }( T=0,  \{ \theta_i \} ).
\end{equation}

In the following discussion, we will employ Eq.\ref{main:boundary} to show that the Gibbs states of toric code models in various space dimensions have a non-negative binegativity spectrum.

\subsection{2d toric code}
The 2d toric code is defined on a 2d lattice with qubits living on links, and the Hamiltonian is $\hat{H}= -\lambda_A \sum_s \hat{A}_s  - \lambda_B \sum_p \hat{B}_p $. $\hat{A}_s(=\prod_{i \in s} \hat{X}_i$) is the product of four Pauli-Xs on links emanating from a star (vertex) $s$ and $\hat{B}_p (= \prod_{i \in p} \hat{Z}_i  )$ is the product of four Pauli-Zs on links on the boundary of a plaquette $p$. This model exhibits a topological order at zero temperature, which is robust against local perturbations. At any non-zero temperature, the order is destroyed  due to the proliferation of point-like excitations (by flipping the sign of stabilizers), which renders the Gibbs state short-range entangled\cite{lu2020detecting}.

\begin{figure}
	\centering
	\begin{subfigure}[b]{0.4\textwidth}
		\includegraphics[width=\textwidth]{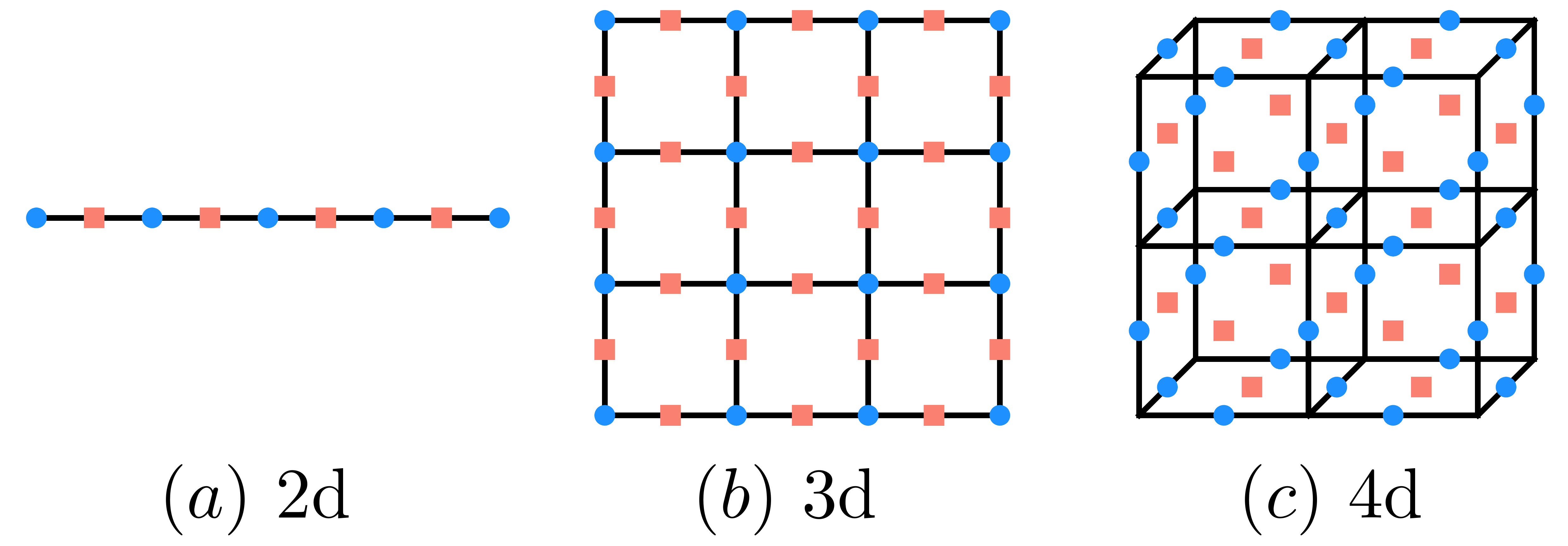}
	\end{subfigure}
	\caption{The boundary stabilizers in toric code for various spatial dimensions, where blue circles and red squares label the X-type stabilizers $A_i$ and Z-type stabilizers $B_j$. (a) 1d bipartition boundary in 2d toric code with $A_i$ defined on sites and $B_j$ defined on links. (b) 2d bipartition boundary in 3d toric code with $A_i$ defined on sites and $B_j$ defined on links.(d) 3d bipartition boundary in 4d toric code with $A_i$ defined on links and $B_j$ defined on faces.
	}    \label{fig:boundary}
\end{figure}

As indicated by Eq.\ref{main:boundary}, the sign of the binegativity spectrum for the Gibbs state is determined by the negativity spectrum for the boundary part of the Gibbs state. It is therefore sufficient to consider a 1d bipartition boundary of size $L$ that involves alternating star and plaquette operators denoted as $\hat{A}_1, \hat{B}_1, \hat{A}_2, \hat{B}_2, \cdots , \hat{A}_L, \hat{B}_L$ (see Fig.\ref{fig:boundary}(a)), with the corresponding boundary Gibbs state $\hat{\rho}_{\partial}\sim e^{ \beta\lambda_A \sum_{i=1}^L \hat{A}_i  + \beta \lambda_B \sum_{i=1}^L \hat{B}_i   }   $. As shown in Ref.\cite{lu_vijay_spt}, the negativity spectrum is given by the correlation functions in the 1d Ising model with onsite fields

\begin{equation}
\begin{split}
&\rho_{\partial}^{\Gamma } (\{  A_i,B_{i} \}) \\
& \sim \sum_{\{  \tau_i \}} \left[\prod_{i} \tau_i^{ \frac{1-A_i}{2}} \right]e^{  -K_A\sum_{i=1}^L\frac{1 -\tau_i }{2}  + \beta \lambda_B  \sum_{ i=1   }^L B_{i} \tau_i \tau_{i+1}   }
\end{split}
\end{equation}
with the field strength being $K_A \equiv  -  \log\left[   \tanh(\beta \lambda_A)  \right]$ and Ising spins $\tau_i = \pm 1$. Using Eq.\ref{main:boundary}, the binegativity spectrum $\abs{\rho^{\Gamma}   }^{\Gamma} (   \{ a_i,  b_{i} \}  )$ with $a_i, b_i \in \{\pm 1\}$ goes as
\begin{equation}
\sum_{   \{A_i, B_{i }\}  } \abs{ \rho_{\partial}^{\Gamma } (T,\{  a_iA_i,b_i B_{i} \})    }    \rho^{\Gamma }_{\partial} (T=0,\{  A_i,B_{i} \})
\end{equation}

When forbidding one type of excitations in the Gibbs state at all temperatures, i.e. $\lambda_A\to \infty$ or $\lambda_B\to \infty$, we  analytically show that the binegativity spectrum is positive (see Appendix.\ref{appendix:2d}). For the general case where both types of excitations are allowed, i.e. $\lambda_A, \lambda_B = O(1)$, since the negativity spectrum is given by correlation functions of 1d Ising model, each eigenvalue can be efficiently computed using a standard transfer matrix method. Utilizing this method, we numerically compute the binegativity spectrum for a given finite size $L$ (up to $L=6$), and find that  the binegativity spectrum  is non-negative at any $\lambda_A, \lambda_B$, and any temperature.

\subsection{3d toric code}

The 3d toric code is defined on a 3d lattice with spins living on links, and the Hamiltonian is $\hat{H}= -\lambda_A \sum_s \hat{A}_s  - \lambda_B \sum_p \hat{B}_p $. $\hat{A}_s$ is the product of six Pauli-Xs on links emanating from a site $s$ and $\hat{B}_p$ is the product of four  Pauli-Zs on links on the boundary of a plaquette $p$. For simplicity, we impose the periodic boundary condition in both $\hat{x}$ and $\hat{y}$ direction while impose the open boudary condition in $\hat{z}$ direction. We divide the system into two part using a plane with a fixed $\hat{z}$ coordinate and there are $L^2$ $\hat{A}_i$ living on sites and $2L^2$ $\hat{B}_{ij}$ living on links in the 2d bipartition surface (see Fig.\ref{fig:boundary}(b)). The boundary Gibbs states is $\hat{\rho}_{\partial}  \sim e^{ - \beta \hat{H}_{\mA \mB} }$ with the boundary Hamiltonian $\hat{H}_{\mA \mB } =   -\lambda_A \sum_i \hat{A}_i  - \lambda_B  \sum_{\expval{ij}}  \hat{B}_{ij}$. In this case, the negativity spectrum is given by correlation functions in 2d Ising model under on-site fields\cite{lu_vijay_spt} 

\begin{equation}
	\begin{split}
&\rho_{\partial}^{ \Gamma   } (  \{ A_i ,B_{ij} \} ) \\
&   \sim \sum_{ \{ \tau_i  \}  }	\left[ \prod_{i } \tau_i^{\frac{1-A_i}{2}  } \right]e^{ -K_A \sum_{i } \frac{1-\tau_i}{2}  +\beta \lambda_B \sum_{\expval{ij}  }  B_{ij} \tau_{i}  \tau_j    },
\end{split}
\end{equation}
where $A_i \in \{\pm 1\}$ determines the spin insertion and $B_{ij} \in \{\pm 1\}$ determines the sign of coupling between two neighboring spins $\tau_i$ and $\tau_j$. 

Using the knowledge of negativity spectrum with Eq.\ref{main:boundary}, we analytically show that its binegativity spectrum is non-negative at any temperatures when either $\lambda_A\to \infty$ or $\lambda_B \to \infty$ (i.e. one species of excitations is forbidden)(see Appendix.\ref{appendix:3d} for details). For the general case with $\lambda_A, \lambda_B  = O(1)$, we numerically confirm the binegativity spectrum remains non-negative, and therefore, PPT entanglement cost is again simply given by entanglement negativity.

From the perspective of non-trivial quantum phases at finite temperature, the limit $\lambda_A\to \infty $ is particularly interesting as it allows for a robust topological order up to  a  non-zero critical temperature $T_c$ by prohibiting point-like excitations. In particular, such a transition in topological order manifests in the structure of entanglement  negativity between the region $\mA$ and $\mB$ separated by a 2d boundary of size $L\times L$\cite{lu2020detecting,lu_vijay_spt} : 
\begin{equation}
E_N = 	 \alpha  L^2 - E_{N,\text{topo}}. 
\end{equation}
While the area-law coefficient $\alpha$ is non-universal, the quantity $ E_{N,\text{topo}}$, dubbed topological entanglement negativity, characterizes the universal, long-range entanglement that diagnoses the finite-temperature topological order. In this case, $ E_{N,\text{topo}}$ takes the value $\log 2$ for $T < T_c$ and $0$ for $T > T_c$, corresponding to the presence and absence of topological order in a Gibbs state. Therefore, our result on the non-negative binegativity spectrum indicates that such a non-trivial entanglement structure probed  by entanglement negativity is operationally meaningful as PPT entanglement cost.

\subsection{4d toric code}
Finally, we discuss the 4d toric code, which only hosts loop-like excitations so the  topological order exisits below a certain critical temperature\cite{dennis2002} even when both $\lambda_A, \lambda_B=O(1)$. The model is defined on a 4d lattice with spins living on plaquettes (2-cells), and the Hamiltonian is $ \hat{H} =-\lambda_A \sum_l \hat{A}_l  -\lambda_B \sum_c \hat{B}_c$.  $\hat{A}_l$ is the product of six Pauli-Xs on plaquettes adjacent to the link $l$ (1-cell) and $\hat{B}_c$ is the product of six  Pauli-Zs on plaquettes  on the boundary of the cube $c$ (3-cell). We consider a bipartition by fixing one of the four spatial coordinates so the bipartition surface is a 3d lattice, and the boundary interaction is $\hat{H}_{\mA \mB }  =    -\lambda_A \sum_l \hat{A}_l  -\lambda_B  \sum_p \hat{B}_p $ with $\hat{A}_l$ living on links and $\hat{B}_p$ living on plaquettes (see Fig.\ref{fig:boundary}(c)). Given the boundary part of the Gibbs state $\hat{\rho }_{\partial} \sim e^{ -  \beta \hat{H}_{\mA \mB }} $, the negativity spectrum can be written as the correlation functions in a 3d Ising gauge theory coupled to matter fields\cite{lu_vijay_spt}: $\rho_{\partial}^{\Gamma }  ( \{ A_l ,B_{p }   \}  )   $
\begin{equation}
\sim \sum_{ \{\tau_l \} } \left[ \prod_{l} \tau_l^{\frac{1-A_l}{2}}\right] e^{ -K_A \sum_{l} \frac{1-\tau_l}{2}   +  \beta \lambda_B \sum_{ p  } B_{p}  \prod_{ l \in \partial p  }\tau_l     },
\end{equation}
where the Ising spins $\tau_l$ are defined on links. When prohibiting one type of excitations, using a calculation similar to 3d toric code, we analytically prove that the binegativity spectrum of the Gibbs state is non-negative (see Appendix.\ref{appendix:4d} for details), indicating the equivalence between PPT entanglement cost and  entanglement negativity.

\section{Summary and Discussion}
In this work, we explore the entanglement structure of many-body systems via an operational meaningful way from the perspective of quantum information theory. Specifically, by computing the binegativity spectrum, we show that the entanglement cost using PPT operations for preparing a thermal Gibbs state of toric code models is exactly given by entanglement negativity, which provides the first exact result on PPT entanglement cost for topologically ordered states of matter at finite temperature. This is notable since quantifying  entanglement cost in mixed states is typically a challenging task, especially for states relevant to many-body systems.

Two questions naturally arise from our work: (i) Given that the toric code models at all temperatures have non-negative binegativity spectrum, does such a result carry over to Gibbs states of any stabilizer models? (ii) Does the equivalence between entanglement negativity and PPT entanglement cost hold true for the toric code models under weak local perturbation? We leave these questions for future study.

In this work, we did not address another aspect of the operational meaning of entanglement, namely, the distillable entanglement. In this regard, an important question is whether entanglement cost and distillable entanglement can coincide, which indicates the reversibility of entanglement manipulation in quantum states. While the answer is affirmative for pure states, mixed states are typically irreversible. One notable exception is given by the anti-symmetric Werner states whose PPT entanglement cost equals PPT distillable entanglement\cite{audenaert2002entanglement}. It is therefore a natural question in the future to explore the structure of distillable entanglement for states relevant to quantum phases of matter, such as Gibbs states of topological stabilizer  models.

Finally, we note that for states that violate the condition $\abs{\rho^{\Gamma}}^{\Gamma} \geq 0$, in general it is difficult to  analyze their PPT entanglement cost since optimization is required among all PPT-preserving operations. This issue was recently addressed by Wang and Wilde\cite{wang2018exact,wang2020cost}, who proposed ``$\kappa$-entanglement'' as a mixed-state entanglement measure, and proved that it  is exactly equal to the PPT entanglement cost. Notably, it can be efficiently computed using the semi-definite programming without needing to optimize over all possible PPT-preserving operations. Therefore, such a quantity may potentially be  useful in understanding the PPT entanglement cost  for a wide range  of quantum many-body systems.


\acknowledgements{We thank Zhi Li and Tim Hsieh for helpful discussions and feedback. T.-C. Lu thanks Sagar Vijay for collaboration on a related subject and acknowledges the support from Perimeter Institute for Theoretical Physics. H.-H. Lin thanks Yi-Zhuang You and Daniel Arovas for helpful discussions. Research at Perimeter Institute is supported in part by the Government of Canada through the Department of Innovation, Science and Economic Development and by the Province of Ontario through the Ministry of Colleges and Universities.
E.-J Kuo acknowledges funding from AFOSR-MURI FA95501610323. 
}

\bibliography{references}

\newpage 
\appendix

\onecolumngrid


\section{Calculation of binegativity spectrum in stabilizer models}\label{appendix:spectrum}

Here we present a general framework for computing the binegativity spectrum, i.e. the eigenspectrum of $\abs{\rho^{\Gamma}}^{\Gamma}   $ for stabilizer models at finite temperature\cite{lu2019singularity,lu2020detecting,lu_vijay_spt}. Consider a set of commuting operators $\{\hat{\theta}_i\}$, one defines a stabilizer Hamiltonian $\hat{H}  =  -  J\sum_i  \hat{\theta}_i$, and   the corresponding Gibbs state at inverse temperature $\beta$ is 
\begin{equation}
	\hat{\rho}\sim  e^{ -\beta \hat{H} } \sim \prod_{i}  (1+ t \hat{\theta}_i ) = \sum_{  \{s_i\}   }  \prod_i (t\hat{\theta}_i)^{s_i},
\end{equation}
with  $t= \tanh(\beta J)$ and  $s_i \in  \{  0,1 \}$. Now taking a partial transpose for a subregion $A$ gives 
\begin{equation}
	\hat{\rho}^{\Gamma}\sim  \sum_{\{s_i \}}  \psi (  \{s_i\}  ) \prod_i (t\hat{\theta}^{\Gamma}_i)^{s_i},
\end{equation}
where $\hat{\theta}^{\Gamma}_i$ is the partial transpose of the stabilizer $\hat{\theta}_i$, and the sign $\psi(\{s_i \})$ is  determined by the anticommutation relation among $\hat{\theta}_i$ in $\prod_i \hat{\theta}_i^{s_i}$ when restricted on the subregion $\mA$. Specifically, defining  $\hat{\theta}_i|_{\mA}  $ as the operator in $\hat{\theta}_i$ supported non-trivially on $\mA$, we introduce a matrix $C$ whose entry $C_{ij}=0, 1$ for $[ \hat{\theta}_i|_{\mA}, \hat{\theta}_j|_{\mA} ] =0  $ and  $\{ \hat{\theta}_i|_{\mA}, \hat{\theta}_j|_{\mA} \} =1  $ respectively. One finds that $\psi(\{s_i\}  ) = (-1)^{  \sum_{  i<j } s_iC_{ij}s_j    }$. Since $\hat{\theta}_j^{\Gamma}$ commutes with each other, the negativity spectrum $\rho^{\Gamma}$, i.e. the eigenspectrum of $\hat{ \rho}^{\Gamma} $, can be obtained by replacing $\hat{\theta}_i^{\Gamma}$ with $\theta_i \in  \{    1,-1\}$:

\begin{equation}
	\rho^{\Gamma}(T,  \{\theta_i\}  ) \sim  \sum_{  \{s_i\}   }  \psi( \{s_i\}   )  \prod_i (t\theta_i)^{s_i}. 
\end{equation}
Given this result, one can express the partially transposed matrix $\hat{\rho}^{\Gamma} $ in terms of the stabilizers $\{ \hat{\theta}_i  \}  $: 
\begin{equation}
	\hat{\rho}^{\Gamma }  = \sum_{\{\theta_i\}}  \rho^{\Gamma    }(T, \{\theta_i\}   ) \prod_{i} \frac{1+  \theta_i \hat{\theta}_i^{\Gamma}   }{2}. 
\end{equation}
Taking the absolute value of the matrix gives 

\begin{equation}
	\abs{		\hat{\rho}^{\Gamma } } =\sum_{\{\theta_i\}}  \abs{\rho^{\Gamma    }(T, \{\theta_i\}   ) }\prod_{i} \frac{1+  \theta_i \hat{\theta}_i^{\Gamma}   }{2}. 
\end{equation}
We can expand the projector $\prod_i \frac{1+ \theta_i \hat{\theta}^{\Gamma}_i}{2} = \sum_{\{\tau_i\}}  \prod_i  (\theta_i \hat{\theta}_i^{\Gamma} )^{\tau_i}  $ with $\tau_i \in  \{  0,1 \}$. Taking a partial transpose gives 
\begin{equation}
	\abs{		\hat{\rho}^{\Gamma } }^{\Gamma} = \sum_{\{\theta_i\}}  \abs{\rho^{\Gamma    }(T, \{\theta_i\}   ) }   \left[ \sum_{\{\tau_i\}} \psi( \{ \tau_i \})    \prod_i  (\theta_i \hat{\theta}_i )^{\tau_i}      \right]   
\end{equation}
Now the binegativity spectrum  $\abs{		\rho^{\Gamma } }^{\Gamma}(  \{g_i \}    )$, i.e. the spectrum of $\abs{		\hat{\rho}^{\Gamma } }^{\Gamma} $, can be obtained by replacing the stabilizer $\hat{\theta}_i$ with $g_i=  \pm 1$:  

\begin{equation}
	\abs{		\rho^{\Gamma } }^{\Gamma} (  \{ g_i   \}  ) = \sum_{\{\theta_i\}}  \abs{\rho^{\Gamma    }(T, \{\theta_i\}   ) }   \left[ \sum_{\{\tau_i\}} \psi(\{\tau_i \}   )    \prod_i  (\theta_i  g_i)^{\tau_i}      \right]     =  \sum_{\{\theta_i\}}   \abs{\rho^{\Gamma    }(T, \{\theta_i\}   ) } \rho^{\Gamma    }(T=0,  \{ \theta_i g_i \}   ).    
\end{equation}

Alternatively, one can use the fact that  $\sum_{\{\theta_i \}} = \sum_{ \{\theta_i g_i\} }  $ to write the binegativity spectrum as

\begin{equation}
	\abs{\rho^{\Gamma}   }^{\Gamma} ( \{ g_i   \}  )= 	\sum_{\{\theta_i \}  }  \abs{\rho^{\Gamma    }(T,  \{ g_i \theta_i \} )  } \rho^{\Gamma    }( T=0,  \{  \theta_i  \} ).
\end{equation}
Therefore, the negativity spectrum allows us to compute the binegativity spectrum.

\section{Calculation of binegativity spectrum in toric code models}
Here we present details on  the binegativity spectrum of $d$-dim toric code model for $d=2, 3, 4 $ at finite temperature. As discussed in the main text, the sign of binegativity spectrum of Gibbs states for the entire system is solely determined by the binegativity spectrum of the boundary part of Gibbs states. We will only focus on the latter, and for notational simplicity, we will just use $\rho$ instead of $\rho_{\partial}$ to denote it.

\subsection{2d toric code}\label{appendix:2d}
The negativity spectrum of Gibbs state in 2d toric code is given by the correlation functions in the 1d Ising model

\begin{equation}
	\rho^{\Gamma } (\{  A_i,B_{i} \})  \sim \sum_{\{  \tau_i \}} \prod_{i} \tau_i^{ \frac{1-A_i}{2}} e^{  -K_A\sum_{i=1}^L\frac{1 -\tau_i }{2}  + \beta \lambda_B  \sum_{ i=1   }^L B_{i} \tau_i \tau_{i+1}   }
\end{equation}
with $K_A \equiv  -  \log\left[   \tanh(\beta \lambda_A)  \right]$ and $\tau_i = \pm 1$. 

\subsubsection{One type of excitations forbidden}
When one type of excitations in the toric code is forbidden by taking either $\lambda_A \to \infty$ or $\lambda_B\to \infty$, we analytically show that the binegativity spectrum is non-negative. Below we present the derivation as $\lambda_B  \to \infty$, and we note that the result as $\lambda_A\to \infty$ trivially follows due to the duality between two types of excitations.

For $\lambda_B  \to \infty$, only the frustration-free $\tau_i$ spin configurations contribute: 

\begin{equation}
	\rho^{\Gamma } (\{  A_i,B_{i} \})  \sim \sum_{\{  \tau_i \}} \prod_{i} \tau_i^{ \frac{1-A_i}{2}} e^{  -K_A\sum_{i=1}^L\frac{1 -\tau_i }{2}  } \prod_i \delta( B_i \tau_i \tau_{i+1}  = 1 ),
\end{equation}
where the constraint $ \prod_i \delta( B_i \tau_i \tau_{i+1}  = 1 )$ implicitly implies that $\{B_i\}$ should satisfy the condition $\prod_{i=1}^L B_i = 1$. In addition, the frustration-free condition suggests that only two $\{  \tau_i\}$ configurations related by a global $Z_2$ spin flip are allowed so the negativity spectrum reads 

\begin{equation}
	\begin{split}
		\rho^{\Gamma } (\{  A_i,B_{i} \})  & \sim \prod_{i} \tau_i^{ \frac{1-A_i}{2}} e^{  -K_A\sum_{i=1}^L\frac{1 -\tau_i }{2}  } +       \prod_{i} (-\tau_i)^{ \frac{1-A_i}{2}} e^{  -K_A\sum_{i=1}^L\frac{1 + \tau_i }{2}  } \\
		&  =  \prod_{i} \tau_i^{ \frac{1-A_i}{2}}   \left[ e^{  -K_A\sum_{i=1}^L\frac{1 -\tau_i }{2}  }  +  \prod_{i} (-1)^{ \frac{1-A_i}{2}}      e^{  -K_A\sum_{i=1}^L\frac{1 + \tau_i }{2}  }     \right],
	\end{split}
\end{equation}
where $\{ \tau_i\}$ are determined by $\{B_i\}$ so that $ B_i \tau_i \tau_{i+1}  = 1 $.  Using the above negativity spectrum, one can derive the binegativity spectrum

\begin{equation}
	\begin{split}
		\abs{\rho^{\Gamma}   }^{\Gamma} (   \{ a_i,  b_{i} \}  )& = \sum_{   \{A_i, B_{i }\}  } \abs{ \rho^{\Gamma } (\{  a_iA_i,b_i B_{i} \},T)    }    \rho^{\Gamma } (\{  A_i,B_{i} \},T=0)    \\
		&   \sim     \sum_{   \{A_i, B_{i }\}  }   \abs{   e^{  -K_A\sum_{i=1}^L\frac{1 -\tau_i }{2}  }  +  \prod_{i} (-1)^{ \frac{1-a_iA_i}{2}}      e^{  -K_A\sum_{i=1}^L\frac{1 + \tau_i }{2}  }   }\prod_{i} \sigma_i^{ \frac{1-A_i}{2}}   \left[1+  \prod_{i} (-1)^{ \frac{1-A_i}{2}}  \right],
	\end{split}
\end{equation}
where we note that $\{\sigma_{i}  \}$ configuration is determined from the $B_i$ configuration as $\sigma_2= \sigma_1 B_1$, $\sigma_3 = \sigma_1 B_1 B_2, \cdots, \sigma_ i  =\sigma_1 \prod_{j=1}^{i-1} B_j     $. Similarly, $\{\tau_i\}$ satisfies $  b_i B_i \tau_i \tau_{i+1}=  b_i \sigma_{i} \sigma_{i+1} \tau_i \tau_{i+1}  =1$. Since $\{\sigma_i\}$ is fixed  by $\{B_i\}$ (up to a global spin flip), summing over $\{B_i \}$ is equivalent to summing $\{\sigma_i\}$:

\begin{equation}
	\begin{split}
		\abs{\rho^{\Gamma}   }^{\Gamma} (   \{ a_i,  b_{i} \}  )    \sim     \sum_{   \{A_i, \sigma_{i }\}  }   \abs{   e^{  -K_A\sum_{i=1}^L\frac{1 -\tau_i }{2}  }  +  \prod_{i} (-1)^{ \frac{1-a_iA_i}{2}}      e^{  -K_A\sum_{i=1}^L\frac{1 + \tau_i }{2}  }   }\prod_{i} \sigma_i^{ \frac{1-A_i}{2}}   \left[1+  \prod_{i} (-1)^{ \frac{1-A_i}{2}}  \right],
	\end{split}
\end{equation}
Since the term $ \left[1+  \prod_{i} (-1)^{ \frac{1-A_i}{2}}  \right] $ imposes a constraint that $\prod_{i=1}^L A_i =1$, the phase factor $ \prod_{i} (-1)^{ \frac{1-a_iA_i}{2}}  = \prod_{i}(a_iA_i) = \prod_i a_i$, and the binegativity spectrum follows

\begin{equation}
	\begin{split}
		\abs{\rho^{\Gamma}   }^{\Gamma} (   \{ a_i,  b_{i} \}  )    \sim     \sum_{   \{A_i, \sigma_{i }\}  } \abs{   e^{  -K_A\sum_{i=1}^L\frac{1 -\tau_i }{2}  }  + \left(\prod_{ i}a_i \right)     e^{  -K_A\sum_{i=1}^L\frac{1 + \tau_i }{2}  }   }   \prod_{i} \sigma_i^{ \frac{1-A_i}{2}}   \left[1+  \prod_{i} (-1)^{ \frac{1-A_i}{2}}  \right],
	\end{split}
\end{equation}
Since $\abs{   e^{  -K_A\sum_{i=1}^L\frac{1 -\tau_i }{2}  }  +  \left(\prod_{i}a_i  \right)  e^{  -K_A\sum_{i=1}^L\frac{1 + \tau_i }{2}  }   }$ satisfy the $Z_2$ symmetry under $\sigma_i\to  - \sigma_i$, the quantity $\prod_{i} \sigma_i^{ \frac{1-A_i}{2}}  $ must obey the symmetry as well in order to have non-zero contribution after summing over $\{\sigma_{i} \}$. This implies the number of $\sigma_i$ in $\prod_i \sigma_i^{ \frac{1-A_i}{2}  }$ should be even, meaning the condition $\prod_i A_i =1$ has implicitly satisfied, and the projector $\left[1+  \prod_{i} (-1)^{ \frac{1-A_i}{2}}  \right]$ can be removed. Therefore,

\begin{equation}
	\begin{split}
		\abs{\rho^{\Gamma}   }^{\Gamma} (   \{ a_i,  b_{i} \}  ) &   \sim     \sum_{   \{A_i, \sigma_{i }\}  }   \abs{   e^{  -K_A\sum_{i=1}^L\frac{1 -\tau_i }{2}  }  + \left(\prod_{ i}a_i \right)          e^{  -K_A\sum_{i=1}^L\frac{1 + \tau_i }{2}  }   }   \prod_{i} \sigma_i^{ \frac{1-A_i}{2}}    \\
		&=\sum_{   \{ \sigma_{i }\}  } \abs{   e^{  -K_A\sum_{i=1}^L\frac{1 -\tau_i }{2}  }  +   \left(\prod_{ i}a_i \right)        e^{  -K_A\sum_{i=1}^L\frac{1 + \tau_i }{2}  }   }   \prod_{i}   \left(  1+\sigma_i \right) \geq  0.
	\end{split}
\end{equation}

\subsubsection{Both types of excitations allowed}
If one allows both charges at finite temperature (i.e. $\lambda_A, \lambda_B =O(1)$), we are unable to analytically compute the binegativity spectrum. However, since the negativity spectrum is given by the correlation functions in 1d Ising model, we employ a standard transfer matrix method to compute negativity spectrum, from which we numerically verify that the binegativity spectrum remains non-negative all temperatures.

\subsection{3d toric code}\label{appendix:3d}

The negativity spectrum in the 3d toric code is given by the correlation functions of the 2d Ising model, where $\{A_i\}$ and $\{B_{ij}\}$ determine the insertion of $\tau_i$ spin variables and the sign of coupling betwen neighboring spins: 
\begin{equation}
	\rho^{\Gamma } (\{  A_i,B_{ij} \})  \sim \sum_{\{  \tau_i \}} \prod_{i} \tau_i^{ \frac{1-A_i}{2}} e^{ -K_A \sum_{i}\frac{1-\tau_i}{2} +  \beta \lambda_B  \sum_{\expval{ij}   } B_{ij} \tau_i \tau_j   }
\end{equation}
with $K_A= -\log \left[   \tanh(\beta \lambda_A)   \right]$.

\subsubsection{Loop-like excitations forbidden}

Taking the limit $\lambda_B \to \infty$ to forbid loop-like excitations, the negativity spectrum is given by 
\begin{equation}
	\rho^{\Gamma } (\{  A_i,B_{ij} \})  \sim \sum_{\{  \tau_i \}} \prod_{i} \tau_i^{ \frac{1-A_i}{2}} e^{-K_A \sum_{i}\frac{1-\tau_i}{2}} \prod_{\expval{ij}} \delta( B_{ij} \tau_i \tau_j=1 ). 
\end{equation}
where the delta function constraint means only the frustration-free $\tau_i$ spin configurations will be allowed. As a result, only two $\{  \tau_i\}$ configurations related by a global $Z_2$ spin flip need to be considered, giving the  negativity spectrum 

\begin{equation}
	\begin{split}
		\rho^{\Gamma } (\{  A_i,B_{ij} \})  & \sim \prod_{i} \tau_i^{ \frac{1-A_i}{2}} e^{  -K_A\sum_{i=1}^L\frac{1 -\tau_i }{2}  } +       \prod_{i} (-\tau_i)^{ \frac{1-A_i}{2}} e^{  -K_A\sum_{i=1}^L\frac{1 + \tau_i }{2}  } \\
		&  =  \prod_{i} \tau_i^{ \frac{1-A_i}{2}}   \left[ e^{  -K_A\sum_{i=1}^L\frac{1 -\tau_i }{2}  }  +  \prod_{i} (-1)^{ \frac{1-A_i}{2}}      e^{  -K_A\sum_{i=1}^L\frac{1 + \tau_i }{2}  }     \right]
	\end{split}
\end{equation}
One notices that the negativity spectrum is exactly the same as the one in the 2d toric code when forbidding one type of charges. As a result, the 3d toric code with loop-like excitations has non-negative binegativity spectrum.

\subsubsection{Point-like excitations forbidden}
Taking the limit $\lambda_A \to \infty$ to forbid point-like charges, the negativity spectrum is given by 
\begin{equation}
	\rho^{\Gamma } (\{  A_i,B_{ij} \},T)  \sim \sum_{\{  \tau_i \}} \prod_{i} \tau_i^{ \frac{1-A_i}{2}} e^{\beta \lambda_B  \sum_{\expval{ij}   } B_{ij} \tau_i \tau_j   }
\end{equation}

It follows that the benegativity spectrum is

\begin{equation}
	\begin{split}
		\abs{\rho^{\Gamma}   }^{\Gamma} (   \{ a_i,  b_{ij} \}  )& = \sum_{   \{A_i, B_{ij }\}  } \abs{ \rho^{\Gamma } (\{  a_i A_i,b_{ij}B_{ij} \},T)    }   \rho^{\Gamma } (\{  A_i,B_{ij} \},T=0)    \\
		&   \sim    \sum_{   \{A_i, B_{ij }\}  }  \abs{   \sum_{\{  \tau_i \}} \prod_{i} \tau_i^{ \frac{1-a_iA_i}{2}} e^{\beta \lambda_B  \sum_{\expval{ij}   }b_{ij} B_{ij} \tau_i \tau_j   }  } \sum_{ \{   \sigma_i\}  }  \prod_{i} \sigma_i^{ \frac{1-A_i}{2}}  \prod_{\expval{ij}}   \delta( B_{ij} \sigma_i  \sigma_j  =1)
	\end{split}
\end{equation}
where $ \delta( B_{ij} \sigma_i  \sigma_j =1  )= \frac{  1+B_{ij} \sigma_i \sigma_j  }{2}$. Note that such a constraint implicitly implies that the product of four $B_{ij}$ on the boundary of a plaquette is one, i.e. $\prod_{\expval{ij} \in \partial p } B_{ij }=1 $, and therefore one can write $B_{ij}=g_ig_j $ by introducing the dual variable $g_i$ living on sites. Consequently,

\begin{equation}
	\begin{split}
		\abs{\rho^{\Gamma}   }^{\Gamma} (   \{ a_i,  b_{ij} \}  )   & \sim    \sum_{   \{A_i, g_i\}  }   \abs{   \sum_{\{  \tau_i \}} \prod_{i} \tau_i^{ \frac{1-a_iA_i}{2}} e^{\beta \lambda_B  \sum_{\expval{ij}   } b_{ij}g_i g_j \tau_i \tau_j   } } \sum_{ \{   \sigma_i\}  }  \prod_{i} \sigma_i^{ \frac{1-A_i}{2}}  \prod_{\expval{ij}}   \delta(g_i g_j  \sigma_i  \sigma_j  =1)\\
		&=		\sum_{   \{A_i \} }   \abs{ \sum_{\{  \tau_i \}} \prod_{i} \tau_i^{ \frac{1-a_iA_i}{2}} e^{\beta \lambda_B  \sum_{\expval{ij}   } b_{ij }\tau_i \tau_j   } }  \sum_{  \{g_i \} }   \sum_{ \{   \sigma_i\}  }  \prod_{i} \sigma_i^{ \frac{1-A_i}{2}}  \prod_{\expval{ij}}   \delta(g_i g_j  \sigma_i  \sigma_j  =1).
	\end{split}
\end{equation}
Now we analyze the term $ \sum_{  \{g_i \} }   \sum_{ \{   \sigma_i\}  }  \prod_{i} \sigma_i^{ \frac{1-A_i}{2}}  \prod_{\expval{ij}}   \delta(g_i g_j  \sigma_i  \sigma_j  =1)$. For a given fixed $\{g_i\}$ configuration, there are two allowed $\{  \sigma_i \}$ configurations related by a global spin flip, and therefore $  \sum_{ \{   \sigma_i\}  }  \prod_{i} \sigma_i^{ \frac{1-A_i}{2}}  \prod_{\expval{ij}}   \delta(g_i g_j  \sigma_i  \sigma_j  =1)   =     \left[ \prod_{i} \sigma_i^{ \frac{1-A_i}{2}}  + \prod_{i} (-\sigma_i)^{ \frac{1-A_i}{2}}    \right]  \prod_{\expval{ij}}   \delta(g_i g_j  \sigma_i  \sigma_j  =1)  =  \prod_{i} \sigma_i^{ \frac{1-A_i}{2}}  ( 1 + \prod_i  A_i )  \prod_{\expval{ij}}   \delta(g_i g_j  \sigma_i  \sigma_j  =1)    $. As $\prod_i A_i$ is fixed at one, $\prod_i  \sigma_i^{\frac{1-A_i}{2}}$ must involve even number of spins, and via the constraint  $\prod_{\expval{ij}}   \delta(g_i g_j  \sigma_i  \sigma_j  =1)  $, we can replace $\sigma_i $ by $g_i$. As a result,

\begin{equation}
	 \sum_{  \{g_i \} }   \sum_{ \{   \sigma_i\}  }  \prod_{i} \sigma_i^{ \frac{1-A_i}{2}}  \prod_{\expval{ij}}   \delta(g_i g_j  \sigma_i  \sigma_j  =1) = 	 \sum_{  \{g_i \} } \prod_i  g_i^{\frac{1-A_i}{2}}  ( 1+\prod_i A_i ) = \prod_i (1+  (-1)^{\frac{1-A_i }{2}} )  ( 1+\prod_i A_i )  \sim  \prod_i (1+ A_i),
\end{equation}
which is simply a projector to enforce $A_i =1$ for all $i$. It follows that

\begin{equation}
	\begin{split}
		\abs{\rho^{\Gamma}   }^{\Gamma} (   \{ a_i,  b_{ij} \}  )  =   \abs{ \sum_{\{  \tau_i \}} \prod_{i} \tau_i^{ \frac{1-a_i}{2}} e^{\beta \lambda_B  \sum_{\expval{ij}   } b_{ij }\tau_i \tau_j   } }  \geq 0.
	\end{split}
\end{equation}

\subsection{4d toric code}\label{appendix:4d}
In the 4d toric code, the boundary stabilizers are $A_l$ and $B_p$ operators living on links and plaquettes in 3d lattice, and the negativity spectrum is given by the Wilson loops of $\tau_l$ spins living on links in the 3d classical Ising gauge theory coupled to matter fields: 

\begin{equation}
	\rho^{\Gamma } (\{  A_l,B_{p} \},T)  \sim \sum_{\{  \tau_l  \}} \prod_{l} \tau_l^{ \frac{1-A_l}{2}} e^{ -K_A \sum_l \frac{1 - \tau_l}{2} +    \beta \lambda_B  \sum_{p}  B_p \prod_{l \in \partial p} \tau_l   }
\end{equation}
with $K_A = -\log[  \tanh(\beta \lambda_A)   ]$, and $\{A_l\}$ and $\{B_p\}$ determining spin insertions and the sign of coupling between $\tau_l $ spins.

Below we show that the binegativity spectrum of the 4d toric code is non-negative when one type of excitations is forbidden at all temperatures (i.e. either $\beta \lambda_A\to \infty$ or $\beta \lambda_B \to \infty$). We present the calculation as $\beta \lambda_A\to \infty$, and we note that the result as $\beta \lambda_B\to \infty$ trivially follows due to the duality between two types of excitations.

As $\beta \lambda_A\to \infty$, i.e. $K_A =0 $,  the matter fields are absent, giving the negativity spectrum:  

\begin{equation}
	\rho^{\Gamma } (\{  A_l,B_{p} \},T)  \sim \sum_{\{  \tau_l  \}} \prod_{l} \tau_l^{ \frac{1-A_l}{2}} e^{\beta \lambda_B  \sum_{p}  B_p \prod_{l \in \partial p} \tau_l   }.
\end{equation}
It follows that the benegativity spectrum is

\begin{equation}
	\begin{split}
		\abs{\rho^{\Gamma}   }^{\Gamma} (   \{ a_l,  b_{p} \}  )& = \sum_{   \{A_l, B_p   \}  } \abs{ \rho^{\Gamma } (\{  a_lA_l,b_p B_p \},T)    }   \rho^{\Gamma } (\{  A_l, B_{p} \},T=0)    \\
		&   \sim    \sum_{   \{A_l, B_{p }\}  }  \abs{   \sum_{\{  \tau_l \}} \prod_{l} \tau_l^{ \frac{1-a_lA_l}{2}} e^{\beta \lambda_B  \sum_{p  } b_pB_p \prod_{l \in \partial p}  \tau_l    }  } \sum_{ \{   \sigma_l\}  }  \prod_{l} \sigma_l^{ \frac{1-A_l}{2}}  \prod_{p}   \delta(B_{p} \prod_{l\in \partial p  } \sigma_l =1)
	\end{split}
\end{equation}
where $ \delta( B_{p}  \prod_{l\in \partial p   } \sigma_l  =1  )= \frac{  1+B_{p}  \prod_{ l \in \partial p} \sigma_l  }{2}$. Note that such a constraint implicitly implies that the product of six $B_p$ on the boundary of the cube $c$ is an identity, and therefore one can write $B_p= \prod_{l\in \partial p} g_l $ by introducing the dual variable $g_l$ defined on links. Consequently,

\begin{equation}
\begin{split}
\abs{\rho^{\Gamma}   }^{\Gamma} (   \{ a_l,  b_{p} \}  )
&   \sim    \sum_{   \{A_l,g_l\}  }  \abs{   \sum_{\{  \tau_l \}} \prod_{l} \tau_l^{ \frac{1-a_lA_l}{2}} e^{\beta \lambda_B  \sum_{p  }   \prod_{l \in \partial p}  (g_l\tau_l   ) }  } \sum_{ \{   \sigma_l\}  }  \prod_{l} \sigma_l^{ \frac{1-A_l}{2}}  \prod_{p}   \delta( \prod_{l\in \partial p  } ( g_l \sigma_l) =1)\\
&\sim    \sum_{   \{A_l \}}  \abs{ \sum_{\{  \tau_l \}} \prod_{l} \tau_l^{ \frac{1- a_lA_l}{2}} e^{\beta \lambda_B  \sum_{p  }   \prod_{l \in \partial p}  \tau_l    }  }    \sum_{ \{  g_l\}  }     \sum_{ \{   \sigma_l\}  }  \prod_{l} \sigma_l^{ \frac{1-A_l}{2}}  \prod_{p}   \delta( \prod_{l\in \partial p  } ( g_l \sigma_l) =1)
\end{split}
\end{equation}

Now we analyze the following term 
\begin{equation}
\begin{split}
   \sum_{ \{  g_l\}  }     \sum_{ \{   \sigma_l\}  }  \prod_{l} \sigma_l^{ \frac{1-A_l}{2}}  \prod_{p}   \delta( \prod_{l\in \partial p  } ( g_l \sigma_l) =1)	 &  =    \sum_{ \{  g_l\}  }     \sum_{ \{   \sigma_l\}  }  \prod_{l} (g_l\sigma_l)^{ \frac{1-A_l}{2}}  \prod_{p}   \delta( \prod_{l\in \partial p  } \sigma_l  =1)	\\
    &  =     \sum_{ \{   \sigma_l\}  }  \prod_{l} \sigma_l^{ \frac{1-A_l}{2}}     \prod_{p}   \delta( \prod_{l\in \partial p  } \sigma_l  =1)	 \prod_{ l }  (1+ A_l)   \\
      &  =     \sum_{ \{   \sigma_l\}  }     \prod_{p}   \delta( \prod_{l\in \partial p  } \sigma_l  =1)	 \prod_{ l }  (1+ A_l). 
\end{split}		
\end{equation}
Using this result,  the binegativity spectrum reads

\begin{equation}
	\begin{split}
		\abs{\rho^{\Gamma}   }^{\Gamma} (   \{ a_l,  b_{p} \}  )
		&\sim   \abs{ \sum_{\{  \tau_l \}} \prod_{l} \tau_l^{ \frac{1-a_l}{2}} e^{\beta \lambda_B  \sum_{p  }   \prod_{l \in \partial p}  \tau_l    }  }        \sum_{ \{   \sigma_l\}  }     \prod_{p}   \delta( \prod_{l\in \partial p  } \sigma_l  =1)	 \prod_{ l }  (1+ A_l)  \geq 0.
	\end{split}
\end{equation}
Therefore the 4d toric code with one type of excitations forbidden has a non-negative binegativity spectrum at any temperatures.

\end{document}